\documentclass[referee]{mnras}

\usepackage{amsmath}
\usepackage{mathtools}
\usepackage{pgf,tikz}
\usepackage{graphicx}

\usepackage[T1]{fontenc}
\usepackage{ae,aecompl}

\usepackage{graphicx}	
\usepackage{amsmath}	
\usepackage{amssymb}	

\title{Modelling solar coronal magnetic fields with physics-informed neural networks}

\author[H. Baty et al.]{
H. Baty,$^{1}$\thanks{E-mail: hubert.baty@unistra.fr}
and V.Vigon$^{2}$
\\
$^{1}$ Observatoire Astronomique de Strasbourg, Université de Strasbourg, CNRS UMR 7550, F-67000 Strasbourg, France\\
$^{2}$ IRMA and INRIA (TONUS team), Université de Strasbourg, 7 Rue René Descartes, 67000 Strasbourg, France\\
}

\begin{document}
\label{firstpage}
\pagerange{\pageref{firstpage}--\pageref{lastpage}}
\maketitle

\begin{abstract}
We present a novel numerical approach aiming at computing equilibria and dynamics structures
 of magnetized plasmas in coronal environments. A technique based on the use of neural networks that integrates 
 the partial differential equations of the model, and called Physics-Informed
 Neural Networks (PINNs), is introduced. The functionality of PINNs is explored via calculation of different 
 magnetohydrodynamic (MHD) equilibrium configurations,
 and also obtention of exact two-dimensional steady-state magnetic reconnection solutions~\cite{cra95}.
 Advantages and drawbacks of PINNs compared to traditional numerical codes are discussed in order to propose future improvements. Interestingly,
 PINNs is a meshfree method in which the obtained solution and associated different order derivatives are quasi-instantaneously generated at any point of the spatial domain. 
 We believe that our results can help to pave the way for future developments of time dependent MHD codes based on PINNs.
\end{abstract}

\begin{keywords}
magnetic fields - magnetic reconnection - MHD - sun: solar corona - neural networks - physics-informed neural networks.
\end{keywords}



\section{Introduction}

Deep learning techniques based on multilayered neural networks (NNs) are actually increasingly used to solve problems in a variety of domains including
computer vision, language processing, game theory, etc.~\cite{lec15}. The idea to use NNs to solve non-linear differential equations is not
new, since it was initially introduced more than $25$ years ago~\cite{lag98}.
This was made popular only recently, following the work of \textcolor {blue} {Raissi et al. (2019)}.
where the class of Physics-Informed Neural Networks (PINNs) application was introduced. Indeed, PINNs 
benefited from technical progress on automatic differentiation and the facilitated use of Python open source software libraries like Tensorflow or Pytorch.

To date, PINNs are already used for many applications like, fluid dynamics~\cite{cai21},
radiative transfer~\cite{mis23}, astrophysics~\cite{bat23,urb23},
and many other ones. The specificity of the PINNs technique is to minimize the equation's residual at some predefined
set of data called collocation points, where the predicted solution must thus ensure the differential equation. To this purpose, a physics-based loss function
associated to the residual is defined and then used. In the original method proposed by \textcolor {blue} {Raissi et al. (2019)}, that is sometimes called vanilla-PINNs
in the literature, the initial/boundary
conditions required to solve the equations are imposed via a second set of data called training points where the solution is known or assumed. The latter constraints
are applied by minimizing a second loss function that is a measure of the error (e.g the mean squared error), i.e. the difference between the predicted solution
and the values imposed by the initial/boundary conditions.
The combination of the two loss functions allows to form a total loss function that is finally used in a gradient descent algorithm.
PINNs does not require a large amount of training data as the sole knowledge of solution at boundary is required for vanilla-PINNs. Note that,
as initially proposed by \textcolor {blue} {Lagaris (1998)}, it is also possible to exactly enforce the boundary conditions
in order to avoid the use of training data set~\cite{urb23}. This consists in forcing the neural networks to always
assign the prescribed value at the boundary by employing a well behaved trial function. For example, when this value is zero (homogeneous
Dirichlet condition), the initial output of the neural network is multiplied by a function that cancels out on the boundary. However, when the boundary conditions are
not homogeneous or the geometry is complex, this technique becomes complicated to implement. For simplicity, we make the choice to apply the vanilla-PINNs variant in this work.

 The aim of this work consists in assessing the advantages and drawbacks of PINNs
 to solve the dynamics of plasmas immersed in the magnetic field of the solar corona.
  To the best of our knowledge, PINNs technique has never been applied
 to such context in astrophysics, at the exceptions of structure of force-free neutron star magnetospheres~\cite{urb23} and for
 probing the solar coronal magnetic field from observations data~\cite{jar23}. 
 {However, similar PINNs techniques have been recently developed for applications to laboratory plasmas. In particular,
 there is a surge of interest for computing MHD equilibria relevant to toroidal magnetic confinement configurations (e.g. tokamaks)
 for which Grad-Shafranov like equations need to be solved~\cite{kal22}. In this work, 
 the functionality of PINNs is explored through application to 
 two particular solar problems. First, we consider the computation of two-dimensional (2D) force-free magnetic equilibria representative of
arcades and loop like structures in the solar corona by solving an associated Grad-Shafranov like equation. Second, our method is extended to a more
complex system of differential equations that is an incompressible resistive MHD set, with the aim to compute 2D magnetic reconnection solutions. More precisely,
in this work we focus on the reconnective annihilation solutions that are particular exact steady-state solutions obtained in
2D cartesian geometry~\cite{cra95}.

 The paper is organized as follows. In Section 2, we first introduce the basics of PINNs approach for solving partial differential equations (PDEs).
 Section 3 presents the application to the computation of two different examples of 2D MHD equilibria relevant for solar corona.
 In Section 4, a PINNs code with the aim to solve the set of 2D steady-state resistive equations in the framework of incompressible MHD
 is presented. In particular, we assess the applicability of our PINNs solver in retrieving exact analytical solutions~\cite{cra95}.
 Finally, conclusions are drawn in Section 5.

\section{The basics of PINNs}
  
    \begin{figure}
\centering
 \includegraphics[scale=0.20]{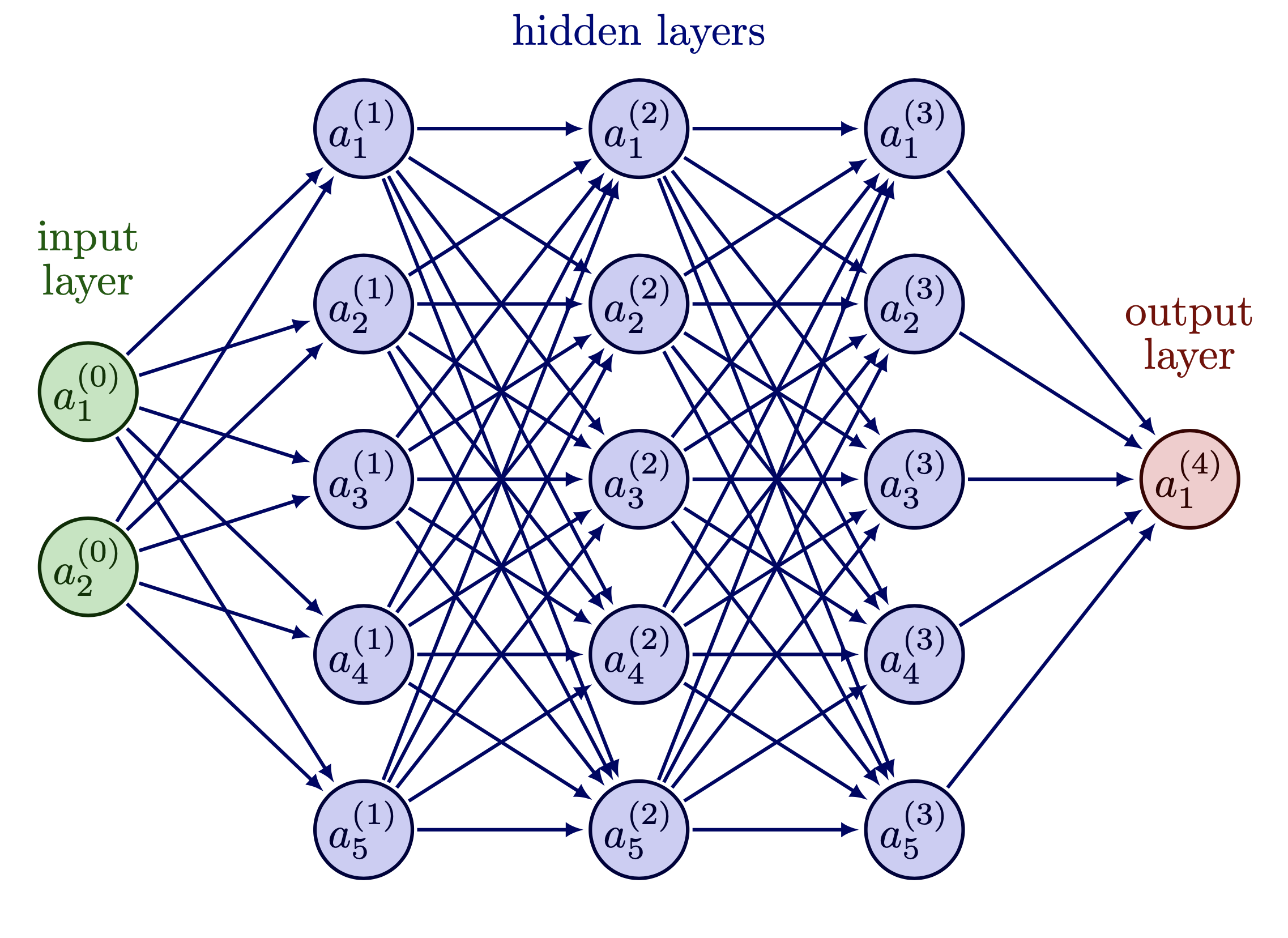}
  \caption{Schematic representation of the structure for a classical neural network having $3$ hidden layers with $5$ neurons per layer, one input layer (two neurons), and
  one output layer (single neuron). The two input neurons represent the spatial coordinates of $\boldsymbol x$ that are fed to the network, and the output neuron is the 
  approximated solution $u_\theta(\boldsymbol x)$.
     }
\label{fig1}
\end{figure}   

   \subsection{The basics of NNs for non linear approximation}
   
In this subsection,  we briefly review how NNs are employed as universal approximators. Let us consider an unknown function $u(\boldsymbol x)$ that could be the solution
of a differential equation, $u_\theta (\boldsymbol {x})$ being its approximated value at given $\boldsymbol x$ value (representing two spatial coordinates) and $\theta$ being a set of model parameters.
  Using a classical feed forward neural network, we can write
  \begin{equation}
u_\theta (\boldsymbol x) =  ( \mathcal{N}^{(L)} \circ \mathcal{N}^{(L-1)} ...\  \mathcal{N}^{(0)}) (\boldsymbol x) ,
\end{equation}
making appear $u_\theta (\bf x)$ as the result of compositions (operator $\circ$ above) of non-linear transformations $\mathcal{N}^{(l)}$ at different layers ($l = 0, 1, ..., L$).
An example of a given feed-forward NN architecture is
schematized in Fig. 1, showing how the neurons for each layer are interconnected. The network is composed of $L+1$ layers including $L-1$ hidden layers
of neurons (e.g. $L = 4$ for Fig. 1). Two neurons are employed for the input layer to represent the two required space coordinates (see below in this paper), and
a single neuron is sufficient to predict the scalar solution $u_\theta (\boldsymbol x)$ in cases involving a single differential equation.
Each transformation can be expressed as
\begin{equation}
 \mathcal{N}^{(l)}  (\boldsymbol x)  =    \sigma ( \boldsymbol{W}^{(l)}  \mathcal{N}^{(l-1)} (\boldsymbol x) +  \boldsymbol{b}^{(l)}) ,
\end{equation}
where we denote the weight matrix and bias vector in the $l$-th layer by $\boldsymbol{W}^{(l)} \in \mathbb{R}^{d_{l-1}  \times d_l}  $  
and $\boldsymbol{b}^{(l)}   \in \mathbb{R}^{d_{l} }$ ($d_l$ being the dimension of the input vector for the $l$-th layer). $\sigma(.)$ is a non linear
activation function, which is applied element-wisely. Such activation function allows the network to map nonlinear relationship that is
fundamental for automatic differentiation and therefore the calculation of the derivatives (see below).
In this work, the most commonly used hyperbolic tangent $tanh$ function is chosen.
Other smooth functions would have led to the same results. However, note that piecewise linear functions ReLU (or Leakly ReLU) would have been
a very bad choice, leading to constant piecewise second derivatives and making impossible to minimize the loss function.
The model is consequently defined by  $\theta =  \lbrace \boldsymbol{W}^{(l)}, \boldsymbol{b}^{(l)}  \rbrace_{l =1,L}$ representing the trainable
parameters of the network. 

The optimization problem aiming to find a non linear approximation $u_\theta (\boldsymbol x) \simeq u (\boldsymbol x)$ is based on the minimization of a function
 $ \mathcal{L}_{data}$, called loss function, that is a measure of the difference between $u_\theta (\boldsymbol  x)$ and  $u (\boldsymbol x)$. In practice, 
 a mean squared error formulation is chosen as
  \begin{equation}
   \mathcal{ L}_{data} (\theta) = \frac  {1} {N_{data} } \sum_{i=1}^{N_{data} } \left|\ u_\theta (\boldsymbol x_i ) - u_i^{data} \right|^2 ,
\end{equation}
where a set of $N_{data}$ data called training data is assumed to be available for $u(\boldsymbol x)$ taken at different $\boldsymbol x_i$ values.
Finally, a gradient descent algorithm is used until convergence towards the minimum is obtained for a predefined accuracy (or a
given maximum iteration number) as
\begin{equation}
             \theta_{k+1} =  \theta_{k} - l_r  \nabla_{\theta}   \mathcal{ L}_{data}  (\theta_k) ,
\end{equation}
for the $k$-th iteration also called epoch in the literature,
leading to 
\begin{equation}
              \theta^{*}  = \operatorname*{argmin}_\theta   \mathcal{ L}_{data} (\theta) ,
\end{equation}
where $l_r$ is known as the learning rate parameter.
This is the so-called training procedure. In this work, we choose the well known $Adam$ optimizer.
The standard automatic differentiation technique is necessary to compute derivatives
(i.e. $\nabla_{\theta}$) with respect to the NN parameters, i.e. weights and biases  \cite{bay18}.
This technique consists in storing the various steps in the the calculation of a compound function, then calculating its gradient using
 the chaine rule. The final goal is to calibrate the trainable parameters $\theta$ (weight matrices and bias vectors) of the network such that $u_\theta (\boldsymbol x)$
approximates the target solution $u(\boldsymbol  x)$. The initialization of the network parameters is done randomly.
The implementation of the algorithm is done using the Tensorflow library,
a classical Python software for machine learning\footnote{https://www.tensorflow.org/}. The gradient descent algorithm is implemented with Keras
using the application programming (API) GradientTape.\footnote{https://keras.io/api/ }
    
     \begin{figure*}
\centering
\includegraphics[scale=0.31]{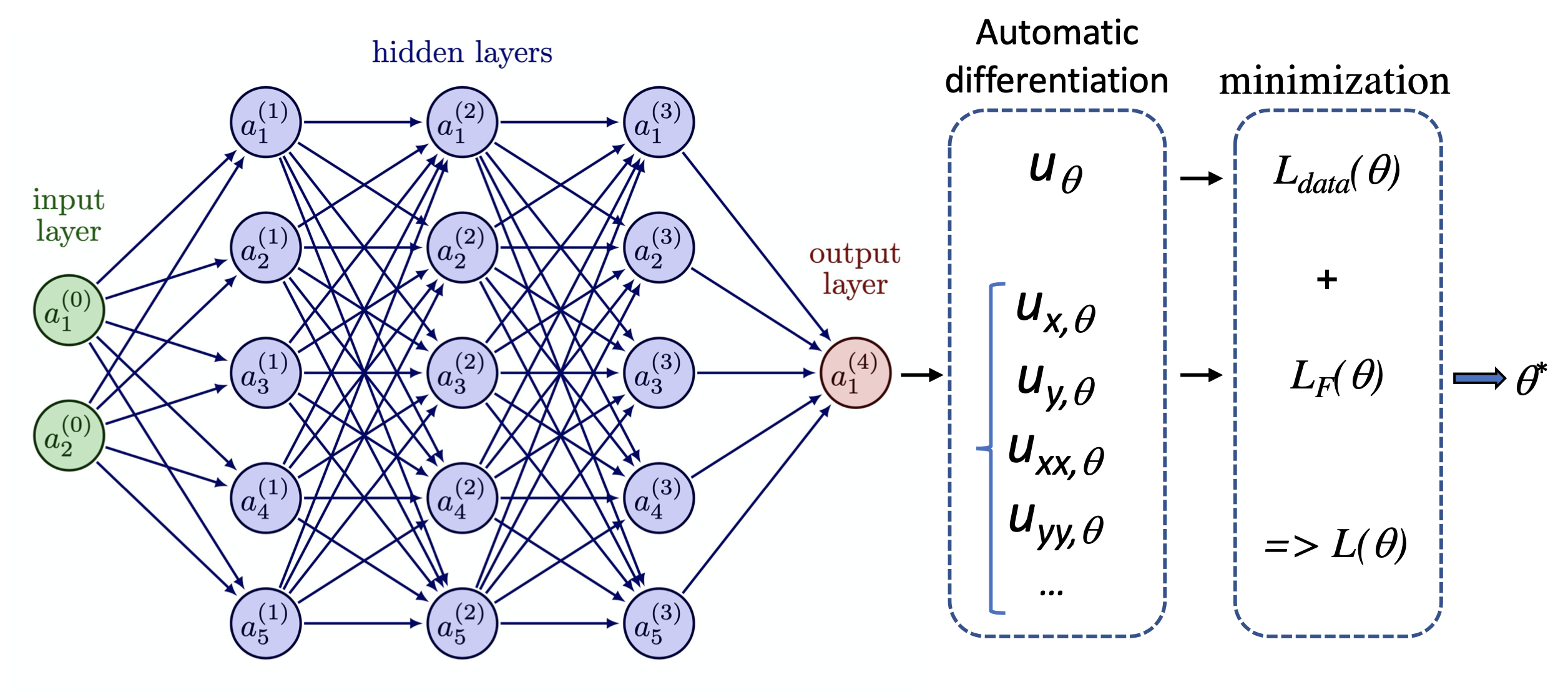}
  \caption{Schematic representation of the network structure PINNs fo solving a single PDE. A NN architecture (see previous figure) is used to evaluate the residual of the equation (via $u_\theta (\boldsymbol x)$
  and associated different partial order derivatives). Two partial loss functions are used to form a total loss function with associated weights (see text) that is finally minimized.   }
\label{fig2}
\end{figure*}   

 \subsection{The basics of PINNs for solving a single PDE}
 
Let consider a function $u(\boldsymbol x)$ satisfying some boundary conditions $u_b(\boldsymbol x)$ at the boundary $ \partial \mathcal{ D}$ of some 2D domain $\mathcal{ D} $.
The previous non linear approximation procedure can be applied once a set of training data is defined at $\boldsymbol x_i$ ($i = 1,..., N_{data})$ where  $u_\theta (\boldsymbol x_i )$ $\simeq u_b(\boldsymbol x_i)$,
and using the minimization of
 \begin{equation}
   \mathcal{ L}_{data} (\theta) = \frac  {1} {N_{data} } \sum_{i=1}^{N_{data} } \left|\ u_\theta (\boldsymbol x_i ) -  u_b(\boldsymbol x_i) \right|^2 .
\end{equation}

In PINNs, the complete minimization is obtained by considering a second loss function that takes into account the equation, so called physics-based loss
function, i.e. $\mathcal{ L}_{ \mathcal{F}}$ hereafter. The latter is defined by using the equation residual that can be written in the simple following form
 \begin{equation}
   \mathcal{F} \left [ u(\boldsymbol x), \boldsymbol x  \right ]  = 0 ,
\end{equation}
where the symbol  $\mathcal{F} $ stands for a nonlinear differential operator.
Indeed, using a second set of data, that are $N_c$ data points located at $\boldsymbol x_j$  ($j \in [1, N_c]$) and
generally called collocation points, we can define the following associated loss function
\begin{equation}
    \mathcal{ L}_{ \mathcal{F}} (\theta) = \frac  {1} {N_c} \sum_{j = 1}^{N_c}  \left| \mathcal{F} [  u_\theta(\boldsymbol x_j), \boldsymbol x_j ]   \right| ^2  ,
\end{equation}
that must be minimized in addition to the training data loss.
As an important property characterizing PINNs, the derivatives of the expected solution with respect to the variable $\boldsymbol x$ (i.e the network input) needed in the previous
loss function are obtained via the automatic differentiation (also used in the gradient descent algorithm described in Section 2.1),
avoiding truncation/discretization errors inevitable in traditional numerical methods.
 In the vanilla-PINN framework, a  total loss function $\mathcal{ L}$ is thus formed as
 \begin{equation}
           \mathcal{ L}  (\theta)  =  \omega_{data}  \mathcal{ L}_{data} (\theta)  +  \omega_{\mathcal{F}}  \mathcal{ L}_{ \mathcal{F}} (\theta),
\end{equation}
where weights $(\omega_{data}, \omega_{\mathcal{F}})$ can be introduced in order to ameliorate the eventual unbalance between
the two partial losses during the training process. These weights and the learning rate can be user-specified or automatically tuned.
In the present work, for simplicity we fix the $\omega_{data} $ and  $\omega_{\mathcal{F}}$ values 
to be constant and equal to unity, and the gradient descent algorithm described in Section 2.1 is thus applied
to the total loss defined in equation 9. A schematic representation summarizing the procedure can be found in Fig. 2.

 \subsection{The basics of PINNs for solving PDEs}

The PINNs solver for a single PDE  can be easily extended for a set of $n$ PDEs
with $m$ desired scalar functions ($n$ being greater or equal to $m$).
Consequently, the output layer must have $m$ neurons instead of one. The training and collocation data
sets must be defined for each function. A physics-based loss function can be defined, that is a weighted sum of 
$n$ physics-based loss functions (one per equation). As a single neural network is used, one must increase the complexity
of the network by increasing the number of neurons and/or the number of hidden layers (see applications in Section 3 and Section 4).

 
 \section{Solving equilibrium equations using PINNs}
 
 Optimization algorithms have been developed for computing MHD equilibria in the solar corona
 using however classical methods where a complex functional is minimized (i.e. without neural networks)~\cite{wie06}.
 Two examples of magnetic solar configurations are considered below that are, an arcade structure, and a curved loop like structure
 obeying a Soloviev Grad-Shafranov equation. Note that, as the exact analytical solutions are known, they are useful in order to evaluate
 the accuracy of the method and also to impose the boundary conditions.
 
  \subsection{Triple arcade structure}
  
       \begin{figure}
\centering
 \includegraphics[scale=0.26]{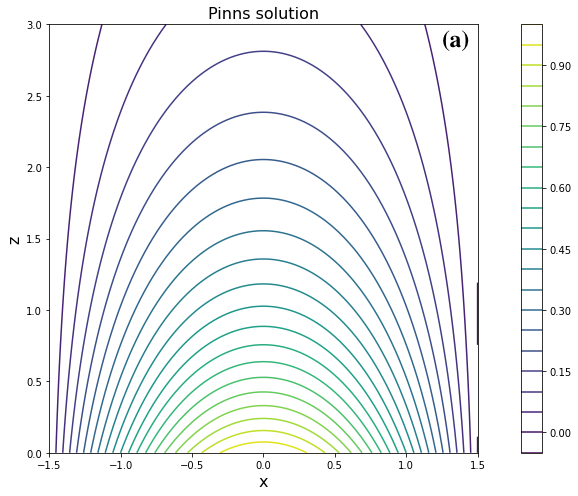}
  \includegraphics[scale=0.26]{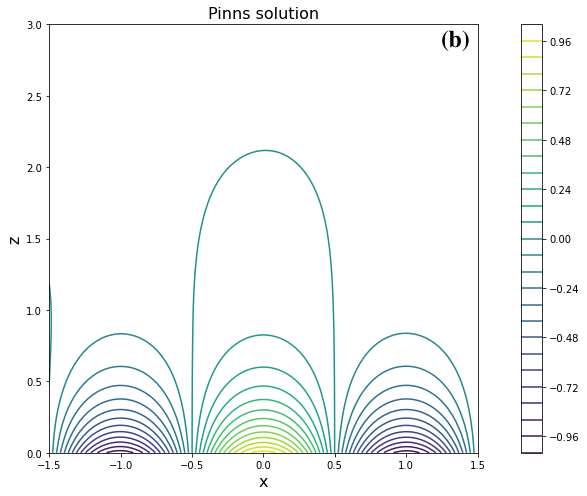}
   \includegraphics[scale=0.26]{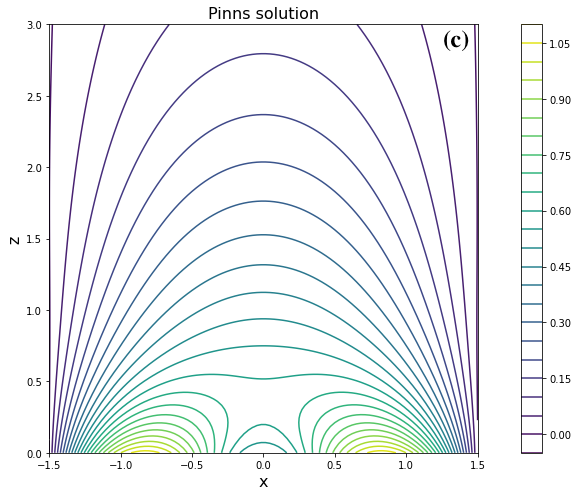}
  \caption{Equilibrium magnetic field lines (iso-contours of $\psi$) obtained with PINNs solver for three particular arcades, (a) dipole-like, (b) quadrupole-like, (c) mixed
  dipole/quadrupole-like configurations (see text).
     }
\label{fig3}
\end{figure}   

 Magnetic arcades are important observed structures in the solar corona \cite{mac99}. Indeed, they are at the heart of solar flares,
 coronal mass ejections (CME), and  other physical processes \cite{jan15,kus04,kuz21,ima13}.
 More precisely, triple arcades are of particular
 importance to explain the initiation of solar flares associated to CME scenario (like the breakout model) in the solar wind \cite{van07}.
  
 Simple force-free models in the framework of two dimensional magnetohydrostatics can be
 deduced from the following equilibrium equation for the scalar field $\psi (x, z)$ representing the $y$ component of the vector potential 
 of the magnetic field in cartesian coordinates \cite{wie98},
     \begin{equation}
       \Delta  \psi + c^2  \psi = 0,
      \end{equation}
where $c$ is a constant and $ \Delta =  \frac {\partial^2} {\partial x^2} +  \frac {\partial^2} {\partial z^2}$ is the cartesian Laplacian operator.
This equation is solved in a spatial domain $(x, z) \in $ $[-L/2 : L/2]  \times [0 : L]$, where $L$ 
is a given reference spatial scale.
This is a linear force-free equilibrium for which the current density and thermal pressure gradient give the
linear form $c^2  \psi $ \cite{wie98}. Exact solutions for triple arcade
structures can be obtained using Fourier-series as
      \begin{equation}
         \psi (x, z) =   \sum_{k=1}^{3}  \exp ( -  \nu_k z)   \left [   a_k \cos  \left ( \frac {k \pi}  {L} x  \right )  \right ] .
      \end{equation}
The latter solution is periodic in $x$, and the relationship $\nu_k^2 =  \frac {k^2 \pi^2}  {L^2} - c^2$ applies as a consequence of equation 10. 

We present the PINNs solutions obtained with $L = 3$, $c = 0.8$, and $a_2 = 0$. Three particular cases are considered below, (a) a dipole-like field
with $a_1 =1$ and $a_3 = 0$, (b) a quadrupole-like field with $a_1 = 0$ and $a_3 = 1$, (c) and a combination of both with
$a_1 = 1$ and  $a_3 = -0.5$. The obtained solutions are plotted in Fig. 3, and can be compared to results previously shown
for a similar set of parameters~\cite{wie98}.

       \begin{figure}
\centering
  \includegraphics[scale=0.44]{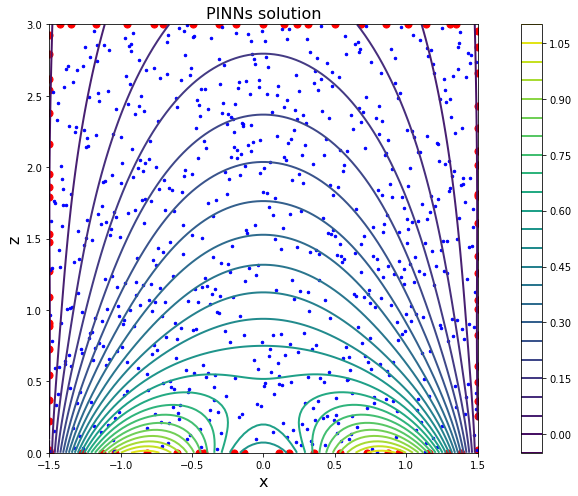}
  \caption{Equilibrium magnetic field lines (iso-contours of $\psi$) obtained with PINNs solver for the mixed
  dipole/quadrupole-like configurations (case c in the previous figure). The spatial location of the training and collocation data sets
  are indicated using red (at boundary layers) and blue dots (inside the domain) respectively.
     }
\label{fig4}
\end{figure}   

       \begin{figure}
\centering
  \includegraphics[scale=0.47]{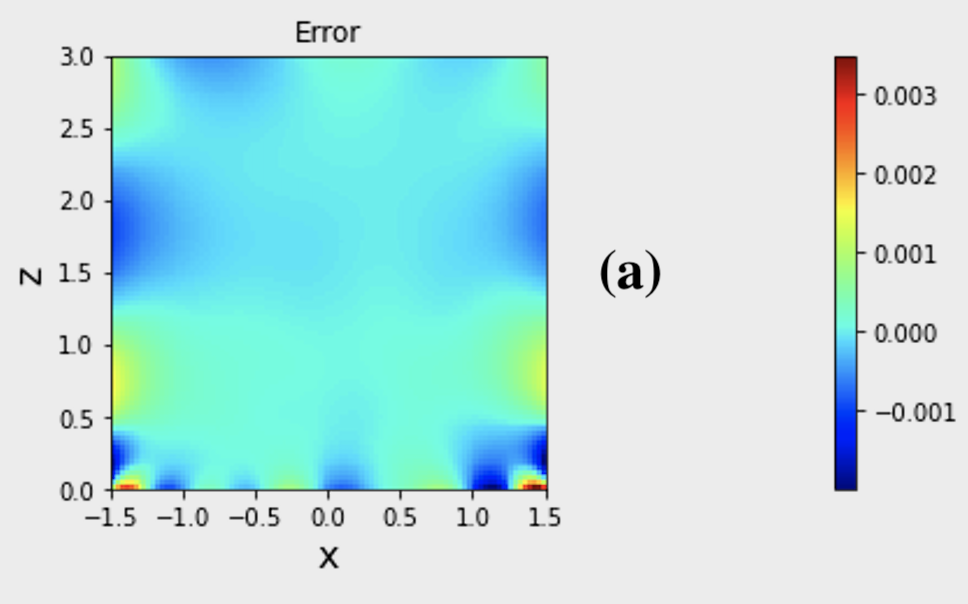}
   \includegraphics[scale=0.50]{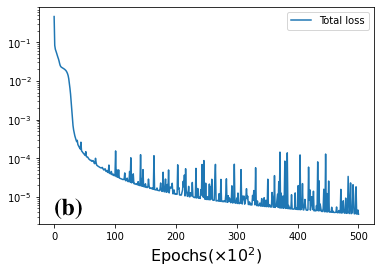}
  \caption{(a) Absolute error distribution (colored iso-contours of the difference between the PINNs and exact solutions) for the arcade case shown in the previous figure.
  (b) Corresponding evolution of the total loss function with the training epochs.
     }
\label{fig5}
\end{figure}   

Moreover, we detailed below the training procedure only for the third case (c), as being similar for the two other cases (a and b).
We have chosen $20$ training data points per boundary layer (i.e. $N_{data} = 80$) with a random distribution, as one can
see in Fig. 4 (with red dots). The exact solution is used to prescribe these training data values.
For the collocation data set, $N_{c} = 700$ points are generated inside the integration domain
using a pseudo-random distribution (latin-hypercube strategy)
as one can see with blue dots. The evolution of the loss function with the training epochs
that is is reported in Fig. 5, shows the convergence toward the predicted solution. Note that the training is stopped after $50000$ epochs
corresponding to a final loss value of order $2 \times 10^{-6} $. We have chosen a network architecture having $7$ hidden layers with $20$ neurons
per layer, and a fixed learning rate of $ l_r = 2 \times 10^{-4} $.
The latter parameters choice slightly influences the results but is not fundamental as long as the number of layers/neurons is not too small~\cite{bat23}.
A faster convergence can be also obtained by taking a variable learning rate with
a decreasing value with the advance of the training process.
The error distribution at the end of the training is plotted in Fig. 5 exhibiting a maximum absolute error of order $0.003$, which also
corresponds to a similar maximum relative error of order $0.003$ (the maximum magnitude solution value being of order unity).
Note that the predicted PINNs solution and associated error distribution are obtained using a third set of points (different from the collocation points)
that is taken to be a uniform grid of $100 \times 100$ points here, otherwise the error could be artificially small (overfitting effect).
One must also note that the error is higher near the boundary due to the higher gradient of the solution and to the coexistence of
data/collocation points in these regions.
In this way, once trained, the network allows to predict the solution quasi-instantaneously
at any point inside the integration domain, without the need for interpolation (as done for 
example with finite-difference methods when the point is situated between two grid points).
The precision of PINNs is known to be very good but less than more traditional methods (like in finite-element codes for example).
This is a general property of minimization techniques based on gradient descent algorithms~\cite{pre07,bat23}.
However, a finer tuning of the network parameters together with the introduction of optimal combinations for weights of the partial
losses can generally ameliorate the results,  which is beyond the scope of the present work.

  \subsection{Grad-Shafranov equilibrium structure: Soloviev solution}
  
          \begin{figure}
\centering
  \includegraphics[scale=0.40]{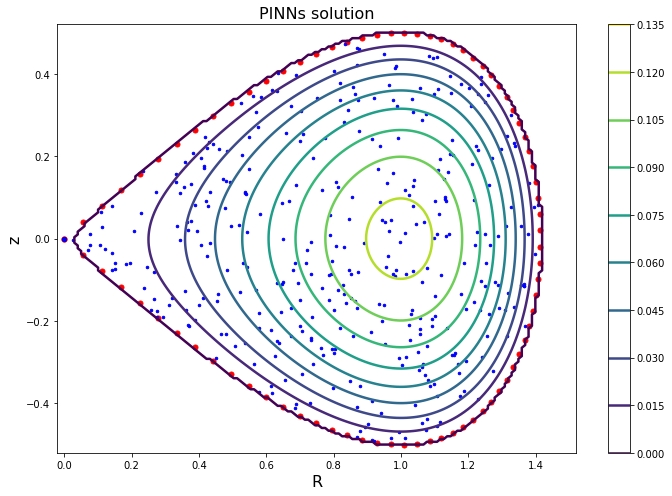}
  \caption{Equilibrium magnetic field lines (iso-contours of $\psi$) obtained with PINNs solver for the Soloviev drop-shaped equilibrium.
  The spatial location of the training and collocation data sets
  are indicated using red (at the boundary layer) and blue dots (inside the domain) respectively.
     }
\label{fig6}
\end{figure}   

       \begin{figure}
\centering
  \includegraphics[scale=0.32]{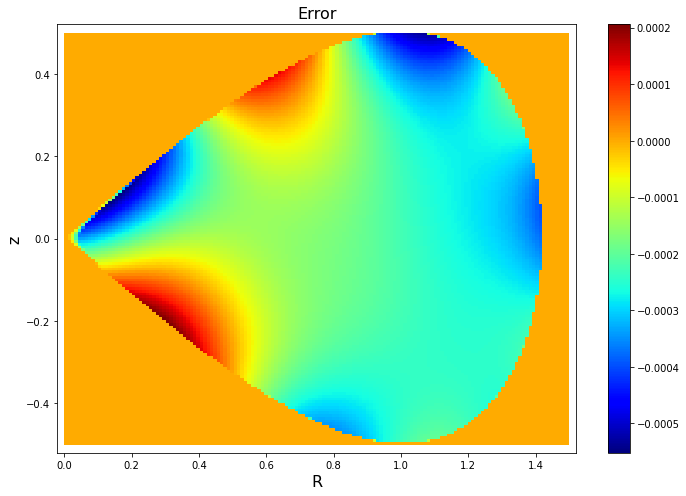}
   \includegraphics[scale=0.52]{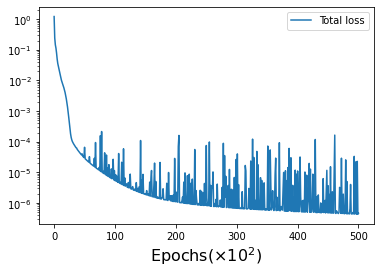}
  \caption{(a) Absolute error distribution (colored iso-contours of the difference between the PINNs and exact solutions) for the drop-shape
   equilibrium shown in the previous figure.
  (b) Evolution of the total loss function with the training epochs for the case shown in the previous figure.
     }
\label{fig7}
\end{figure}   

Equilibrium curved magnetic structures represent another important issue in solar physics. Indeed, the latter obey the
solutions of Grad-Shafranov (GS) equation that is obtained in the axisymmetric approximation. For example, GS equation and its solution are
often used for magnetic clouds reconstruction (e.g. in order to determine their geometries from observations) \cite{isa11}.
GS like solutions are also important to model the CME phenomenon for which a simple force-free spheromak solution is used~\cite{ver19,shi16}. In the latter context, particular solutions
of GS equation called Soloviev solutions can be also implemented as time dependent boundary conditions, leading to
a more realistic and self-consistent CME evolution model and better predictions \cite{lin23}.
 
Following the formulation deduced using ($R, z$) cylindrical like variables in the plane perpendicular to the toroidal angle, the GS equation can be written as
      \begin{equation}
       -   \left [  \frac { \partial^2 \psi  } { \partial R^2 } +  \frac { \partial^2 \psi  } { \partial z^2 }  -  \frac { 1} { R}   \frac { \partial \psi  }  { \partial R }   \right ] = F (R , z,  \psi)
      \end{equation}
where $F$ is a term containing the current density flowing in the toroidal direction~\cite{der11}.
Assuming the particular form for $F$, $F = \alpha R^2 + \beta$ (where $\alpha$ and $\beta$ are constant), allows the obtention of Soloviev solutions~\cite{sol75}.
More precisely, taking $F = f_0 (R^2 + R_0^2)$ leads to the exact solution 
      \begin{equation}
         \psi =  \frac { f_0 R_0^2  } { 2 }  \left [   a^2 - z^2    -    \frac { (R^2 - R_0^2)^2  } { 4 R_0^2 }     \right ]
      \end{equation}
in a spatial domain $\mathcal{ D} $ bounded by its frontier $ \partial \mathcal{ D}$ defined as follows,
    \begin{equation}
  \partial \mathcal{ D} =  \left [ R = R_0 \sqrt   {1 + \frac  { 2 a \cos \alpha  } { R_0 }   } , z = a R_0 \sin \alpha, \alpha= [0: 2 \pi]  \right ] ,
     \end{equation}
     and having a Dirichlet-type boundary condition $\psi = 0$ \cite{der11}. 
 The solution has a drop-shaped structure, that have an $X$-point at $(z = 0, R = 0)$ as $\frac { \partial  \psi} { \partial  z} = \frac { \partial  \psi} { \partial  R} = 0$ at this point.
 Note that similar Soloviev solutions can be also obtained using a different parametrization in order to approximate axisymmetric solutions of tokamak configuration
 having a D-shaped geometry, that are beyond the scope of the present work.

 We present the results obtained with our PINNs solver in Figs 6-7 for finding the solution of equations 12 and 14.
 We have used the following solutions parameter values, $f_0 = 1$, $a=0.5$, 
 and $R_0 = 1$. The network architecture is similar to the arcade case where
 $7$ hidden layers with $20$ neurons per layer were chosen, which consequently represent a number of $2601$ trainable parameters for $\theta$.
We have used $80$ training data points (i.e. $N_{data} = 80$) with a distribution based on a uniform $ \alpha$ angle generator,
 and randomly distributed $N_{c} = 870$ collocation points inside the integration domain. The results are obtained after a training process
 with a maximum of $50000$ epochs. The convergence of the loss function is initially very fast (typically during the first $10000$ epochs) and
 is much more slower after, as already observed previously for the arcade problem.
 When comparing to the exact solution, the relative error of PINNs solver is similar (with a slightly higher value) compared to the arcade problem.
 However, a smaller error is expected with a finer tuning of the different parameters and/or with a longer training procedure.

  \section{Steady-state magnetic reconnection}
  
 Magnetic reconnection plays a fundamental role for release of magnetic energy in solar flares and coronal mass ejections. The mechanism
 has been extensively investigated over the last 50 years ~\cite{pri00} including exact analytical
 solutions for steady state reconnection~\cite{son75,cra95} and numerical
 time dependent reconnection~\cite{bat14,bat19} in the MHD framework approximation.
In incompressible inviscid plasmas, the particular 2D exact solution obtained by  \textcolor {blue} {Craig \& Henton (1995)} that is the generalization of the previously
one introduced by \textcolor {blue} {Sonnerup \& Priest (1975)} is of particular interest in order to test our PINNs solver.

\subsection{Incompressible MHD equations} 
 
 We consider the following set of steady-state incompressible resistive MHD equations written in usual dimensionless units (i.e.
 the magnetic permeability and plasma density are taken to be unity).
 The flow velocity obeys the inviscid equation 
 \begin{equation}
 \boldsymbol V  \cdot \nabla \boldsymbol  V - (\nabla \times \boldsymbol  B)  \times \boldsymbol  {B} + \nabla  P = 0 ,
 \end{equation}
 which is written in a residual form ready to be solved by our PINNs algorithm. The thermal pressure $P$ (via its gradient) is
 necessary to ensure the equilibrium when using the velocity equation.
 The flow velocity vector is also constrained by the incompressibility assumption
  \begin{equation}
\nabla   \cdot  \boldsymbol  V = 0.
 \end{equation}
On the other hand, using the Maxwell-Faraday law and Ohm's law, the magnetic field vector is known to follow the equation
\begin{equation}
    \nabla \times ( \boldsymbol  V   \times \boldsymbol  B )  +  \eta \nabla^2  \boldsymbol  B = 0 ,
 \end{equation}
accompanied by the solenoidal condition
  \begin{equation}
\nabla   \cdot  \boldsymbol  B = 0 .
 \end{equation}
Finally, note that the resistivity $\eta$ is assumed to be uniform in this work.

 \subsection{Magnetic annihilation and reconnective diffusion solutions}  
 
 Magnetic annihilation solution is a particular 2D magnetic reconnection process
in which two anti-parallel regions of magnetic field (directed along the $y$ directions) are swept
together by the incompressible plasma flow and
destroy one another~\cite{son75}. The solution is based on a stagnation-point flow,
\begin{equation}
{\boldsymbol V } =  \left ( - \alpha x, \alpha y \right) ,
 \end{equation}
where $ \alpha$ is a positive real given constant. In the limit of vanishing viscosity, the exact steady state solution for the magnetic field vector is,
\begin{equation}
{\boldsymbol  B } = \left ( 0, B_y (x) \right) ,
 \end{equation}
with
\begin{equation}
B_y (x) = \frac  { E_d}  {\eta \mu   }  Daw(\mu x) ,
 \end{equation}
where $E_d$ is the magnitude of a uniform electric field perpendicular to the $(x, y)$ plane, $\mu ^2 = \alpha / (2 \eta)$
with $\eta$ being the electrical resistivity of the plasma, and
$Daw (x)$ is the Dawson function given by
\begin{equation}
Daw (x) = \int_{0}  ^{x}    \exp (t^2 - x^2)   dt    .
 \end{equation}
The role of $E_d$ is to control the rate of energy conversion.
In the limit of small resistivity $\eta$, this solution exhibits a strong current sheet centered over the stagnation-point flow with a thickness in the $x$-direction proportional to $\eta^{1/2}$.

As an natural extension of the previous reconnection model, the solution of the called reconnective diffusion solution has been
obtained by \textcolor {blue} {Craig \& Henton (1995)}.
It corresponds to the velocity and magnetic field profiles of the form:
\begin{equation}
{\boldsymbol  V } =  \left ( - \alpha x, \alpha y -   \frac  {\beta }   {\alpha }  \frac  {E_d }   {\eta \mu }   Daw(\mu x)   \right) ,
 \end{equation}
and
\begin{equation}
{\boldsymbol  B } = \left ( \beta  x   , - \beta  y  +  \frac  { E_d}  {\eta \mu   }  Daw(\mu x)   \right) ,
 \end{equation}
respectively. The new definition of $ \mu$ parameter is now, 
\begin{equation}
 \mu^2 =   \frac  { \alpha^2 - \beta^2  }  {2 \eta \alpha }  ,
 \end{equation}
 where an additional real parameter $\beta <  \alpha$ is introduced.
Note that the annihilation solution is naturally recovered as a particular case when $ \beta = 0$.
The reconnective diffusion exhibits diffusion across one separatrix like the annihilation solution,
but the dominant process across the other separatrix is advection like in a classical reconnection picture.
As a shear flow exists across a global current layer, there is a symmetry breaking compared
to the annihilation process~\cite{wat98,wat98b,bat16}.

         \begin{figure}
\centering
  \includegraphics[scale=0.4]{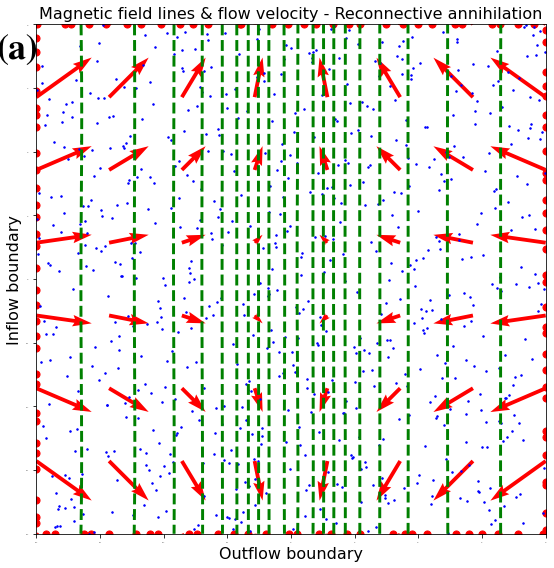}
   \includegraphics[scale=0.4]{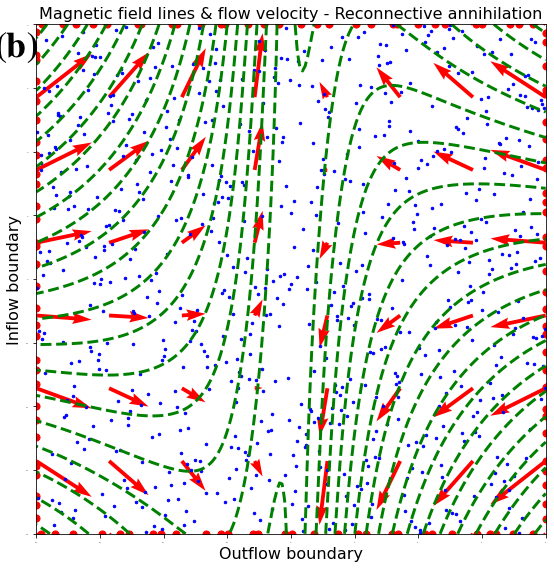}
  \caption{(a) Pure annihilation ($\beta=0$), (b) and reconnective diffusion with $\beta=0.5$ solutions using PINNs solver. Magnetic
  field lines and flow velocity are plotted using iso-contours and red arrows respectively. The location of training and collocation data points are visible
  with red and blue dots respectively.
     }
\label{fig8}
\end{figure}   

 \subsection{Solving steady state magnetic reconnection using PINNs}

Our PINNs solver must therefore treat $6$ scalar equations, that are the two divergence free conditions, two scalar equations for velocity components, and
two scalar equations for magnetic field components, together with the use of $6$ corresponding partial physics-based loss functions. As now $5$ unknown variables
(i.e. $V_x, V_y, B_x, B_y$ and $P$) represent the problem solution, the output layer must at least include $5$ corresponding neurons. In practice, we
have used $5$ neurons, adding a sixth neuron for a magnetic flux function $\psi$ (in order to be used for plotting magnetic field lines) as $B_x = \frac { \partial  \psi} { \partial  y}$
and $B_y = - \frac { \partial  \psi} { \partial  x}$.

Following the same procedure previously used for solving equilibria, the magnetic annihilation and reconnective solutions have been nicely obtained.
Indeed, the results are plotted in Fig. 8 for two values of the $\beta$ parameter, i.e. for $\beta = 0$ thus selecting the pure annihilation solution
and for $\beta = 0.5$ selecting a reconnective diffusion one. The other chosen physical parameters are
$E_d = 0.1$, $\alpha = 1$, and $\eta = 0.01$. The integration is done on a $[-1:1]^2$ square spatial domain.

As concerns the architecture of the network, $9$ hidden layers with $30$ neurons per layer are chosen, which represent
a corresponding number of $7716$ trainable parameters for $\theta$.
We have used $N_{data} = 120$ training data points (i.e. $30$ for each boundary layer) with a random distribution,
 and randomly distributed $N_{c} = 700$ points inside the integration domain. 
 The exact solutions for magnetic field and flow velocity are used to prescribe these training data values.
The results are obtained after a training process with $25000$ epochs employing a learning rate $l_r = 2 \times 10^{-4}$.

The solutions obtained with PINNs solver are compared to the exact analytical ones. The results for $\beta=0.5$ are
plotted in Figs 9-11. A maximum absolute error of order $3 \times 10^{-3} $ is visible on the maps showing the spatial error distribution 
of the magnetic field and velocity flow components, as one can see in Fig. 9 and Fig. 10 respectively. 
Contrary to the previous results obtained for the equilibrium solvers, the error is higher in the central region
due to the higher gradient.
One dimensional cuts 
for different given $x$ and $y$ values plotted in Fig. 11 also confirm the very good precision properties of the solver.
Similar results with similar performances can be also obtained for other $\beta$ values. However, for cases using smaller
resistivity values, the training requires a significantly higher number of collocation points in order to 
resolve the central layer that have a thickness in the $x$-direction proportional to $\eta^{1/2}$.
In practice, it is also possible to use a particular spatial distribution of collocation points having an accumulation in the central region,
that is beyond the scope of this study.

   \begin{figure}
\centering
  \includegraphics[scale=0.47]{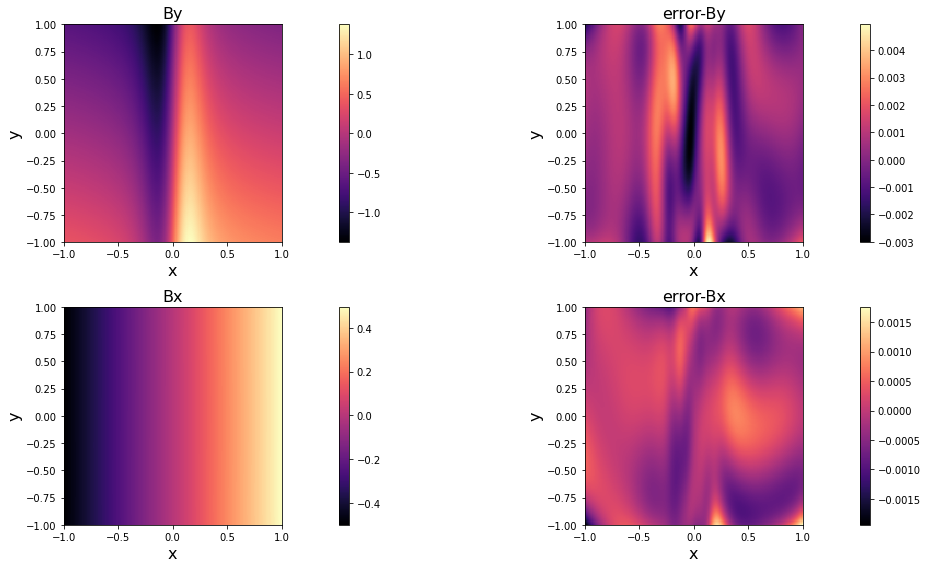}
  \caption{Colored iso-contours of the $B_y$ and $B_x$ magnetic field components predicted by PINNs solver, and associated absolute error
  distributions.
     }
\label{fig9}
\end{figure}   

   \begin{figure}
\centering
  \includegraphics[scale=0.47]{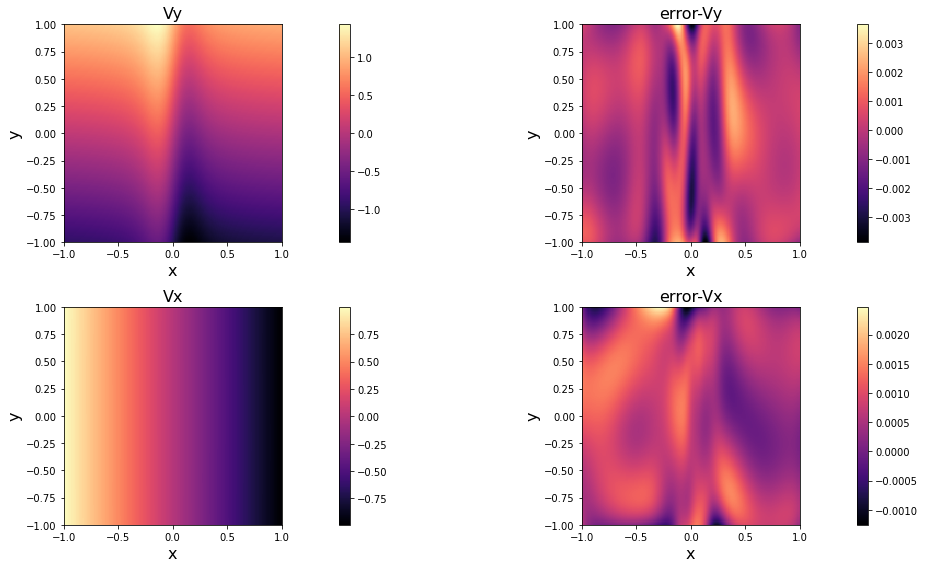}
  \caption{Colored iso-contours of the $V_y$ and $V_x$ velocity flow components predicted by PINNs solver, and associated absolute error
  distributions.
     }
\label{fig10}
\end{figure}   

   \begin{figure}
\centering
  \includegraphics[scale=0.50]{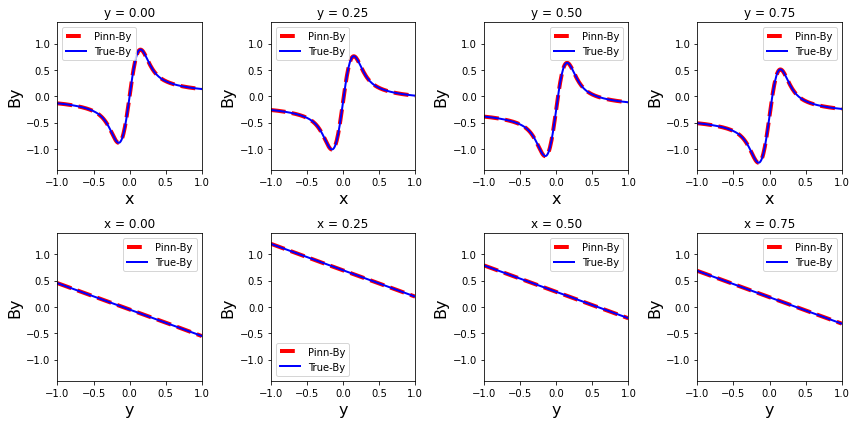}
  \caption{One-dimensional $B_y$ component (red colour) obtained for different $x$ and $y$ particular values (see legend)
  compared with the exact analytical solution (blue colour).
     }
\label{fig11}
\end{figure}

\section{Conclusions}

In this work, we show that PINNs are interesting tools for solving PDEs. In particular, they represent alternatives to traditional/classical
numerical methods for modelling magnetic field dynamics of the solar conona. 
As a first example of application, PINNs-based solvers can easily handle finding equilibrium configurations via solving Grad-shafranov like equations, without the need to
 involve a spatial mesh over which differential operators are discretized in order to solve a large linear system.
 Second, it is shown that PINNs solvers can also offer an alternative to classical MHD codes for modelling dynamics of the solar corona. Indeed, 
exact particular steady state magnetic reconnection solution of 2D incompressible resistive MHD equations is easily recovered in this work~\cite{cra95}.
Compared to traditional numerical methods, they present some advantages listed below.
\begin{enumerate}   
      \item   
      The technique does not require meshing the domain. Indeed, the implementation simply requires
      the use of a a dataset of collocation points arbitrarily chosen inside the domain. It can therefore easily be applied to curvilinear geometries
      or complex domains (for example with holes). The only constraint is to define points in the domain, which is simpler than building meshes.      
      \item  Once trained, the technique makes it possible to calculate the solution at any point of the domain. This allows, for example, to
      zoom in on part of the domain without the need for interpolation. Moreover, this predicted solution is quasi-instantaneously (in a fraction of second) generated, as the latter
       is a function fully determined by the set of parameters $\theta$.         
         \item   The formulation based on the equations residuals (e.g. second order derivative form) does not require the use of some equivalent system
          of first order differential equations. The solution derivative with respect to the spatial variable is also quasi-instantaneously obtained
          with an accuracy similar to the solution.   
     \end{enumerate} 

However, our results also highlight some drawbacks listed below.
      \begin{enumerate}   
        \item    Even if the accuracy obtained in this work is excellent, PINNs seem to be potentially less accurate than classical methods
      where for example refining a grid allows a precision close to the machine one. 
      This limitation is partly inherent to minimization techniques. Nevertheless, our results could be slightly ameliorated (see the second point below).
      
      \item    The training process depends on a combination of many parameters like, the learning rate, the weights in the loss function (not considered in this study), 
     and the architecture of the network, which determines the efficiency (speed and accuracy) of the minimization~\cite{bat23}. Consequently, a finer tuning
     using some adaptive techniques is possible in order to ameliorate the results. However, this is not a simple task that is beyond the scope of
     the present work.
       \end{enumerate}
  
Anyway, PINNs are promising tools that are called upon to develop in future years for the following reasons.
Ameliorations using self-adaptive techniques are expected in order to improve the previously cited drawbacks~\cite{kar21,cuo22}. 
As shown in this work, they also offer a different and complementary approach to traditional methods. Once trained,
the network output being an analytic-like expression (see equations 1-2), 
the solution and derivatives can be quasi-instantaneously generated in the trained spatial domain. Consequently, the solution obtained with our PINNs methods is valid over
the entire domain without the need for spatial interpolation as in classical numerical schemes. 
Another strong promising potentiality offered by PINNs approach is the possibility to learn a family of different solutions with the same neural network~\cite{bat23}.
Indeed, the use of an input layer considering additional variable parameters (it could be the resistivity or/and the $\beta$ parameters
in case of the magnetic reconnection problem) would allow to learn multiple solutions for ranges of variation of these parameters.
We are actually developing such important applications, as this is clearly a particular potentiality of PINNs technique that is not
possible when using traditional numerical schemes.
Finally, another way of using PINNs is to combine a PINNs solver with classical MHD simulations, as this is already under exploitation
for hydrodynamics.
Indeed, data obtained from classical simulations in a first step (e.g. magnetic reconnection ones for different resistivity values)
can be used as extra training data in the neural network training process in order to learn the different associated solutions.
Thus, in the second step, PINNs solver can be used to generate a new solution corresponding to another
parameter value (e.g. resistivity). In other words, PINNs method can serve as a reduced model of a given problem,
avoiding numerous long and costly calculations. 
The computation time needed to obtain the results presented in this work (for a standard single CPU computer) is of order a few minutes
in case of the arcade/equilibrium equations and a few tens of minutes for the reconnection problem.
This is probably faster than obtained with traditional methods on a similar computer. A even faster computation is of course possible when using GPU and multi-GPU.

Beyond the above potentialities, more studies are obviously needed to extend the examples of application presented in this work.
First, the reconstruction of the solar coronal magnetic field in a more realistic three dimensional (3D) geometry could be
a challenging project. The transition to 3D version doesn't necessitate special adaptation (only additional input/output neurons),
but the computation time would be higher as a higher number of points and possibly a larger/deeper neural network are required.

Second, using a PINNs solver for a time dependent MHD dynamics is also actually under development
either for exploitation in combination with a classical MHD code or not.

\section*{Acknowledgements}
The authors thank Emmanuel Franck and Victor Michel-Dansac (IRMA, Strasbourg) for fruitful discussions on PINNs technique. We
also sincerely thank the anonymous referee for useful suggestions that helped improve the content of the paper.

\section*{Data Availability}
Data will be made available on reasonable request to the corresponding author.








\bsp	
\label{lastpage}
\end{document}